\documentclass[a4paper,fleqn,usenatbib]{mnras}
\usepackage[T1]{fontenc}
\usepackage{ae,aecompl}
\usepackage{graphicx}
\usepackage{amsmath}
\usepackage{amssymb}
\usepackage{txfonts}
\usepackage{float}
\usepackage{adjustbox}
\usepackage{siunitx}

\usepackage{tensor}

\def\spirou#1{{\bf}}

\graphicspath{{figures/}}

\newcommand{\ltsima}{$\; \buildrel < \over \sim \;$}
\newcommand{\lsim}{\lower.5ex\hbox{\ltsima}}
\newcommand{\gtsima}{$\; \buildrel > \over \sim \;$}
\newcommand{\gsim}{\lower.5ex\hbox{\gtsima}}
\newcommand{\bra}{\langle}
\newcommand{\ket}{\rangle}

\newcommand{\dd}{\mathrm{d}}

\newcommand{\ci}{\mathrm{i}}

\newcommand{\likeli}{\mathcal{L}}

\title[partition functions and non-Gaussian likelihoods]
{Partition function approach to non-Gaussian likelihoods: Formalism and expansions for weakly non-Gaussian cosmological inference}
\author[L. R{\"o}ver, L.C. Bartels, B.M. Sch{\"a}fer]
{Lennart R{\"o}ver$^{1,2}$\thanks{e-mail: l.roever@thphys.uni-heidelberg.de}, Lea Carlotta Bartels$^2$, Bj{\"o}rn Malte Sch{\"a}fer$^2$\thanks{e-mail: bjoern.malte.schaefer@uni-heidelberg.de}\\
$^1$ Institut f{\"u}r Theoretische Physik, Universit{\"a}t Heidelberg, Philosophenweg 16, 69120 Heidelberg, Germany\\
$^2$Zentrum f{\"u}r Astronomie der Universit{\"a}t Heidelberg, Astronomisches Rechen-Institut, Philosophenweg 12, 69120 Heidelberg, Germany
}

\begin{document}
\pagerange{\pageref{firstpage}--\pageref{lastpage}}
\pubyear{2022}
\maketitle
\label{firstpage}

\begin{abstract}
Non-Gaussian likelihoods, ubiquitous throughout cosmology, are a direct consequence of nonlinearities in the physical model. Their treatment requires Monte-Carlo Markov-chain or more advanced sampling methods for the determination of confidence contours. As an alternative, we construct canonical partition functions as Laplace-transforms of the Bayesian evidence, from which MCMC-methods would sample microstates. Cumulants of order $n$ of the posterior distribution follow by direct $n$-fold differentiation of the logarithmic partition function, recovering the classic Fisher-matrix formalism at second order. We connect this approach for weakly non-Gaussianities to the DALI- and Gram-Charlier expansions and demonstrate the validity with a supernova-likelihood on the cosmological parameters $\Omega_m$ and $w$. We comment on extensions of the canonical partition function to include kinetic energies in order to bridge to Hamilton Monte-Carlo sampling, and on ensemble Markov-chain methods, as they would result from transitioning to macrocanonical partition functions depending on a chemical potential. Lastly we demonstrate the relationship of the partition function approach to the Cram{\'e}r-Rao boundary and to information entropies.
\end{abstract}


\onecolumn

\section{introduction}
In physical systems, Monte-Carlo Markov-chain methods and in particular the the Metropolis-Hastings algorithm \citep{metropolis_monte-carlo:_1985, metropolis_equation_1953} provide an efficient and physically transparent way of sampling microstates from an ensemble, and became indispensable in simulations in statistical physics. They are an efficient numerical tool to compute thermodynamical quantities from a model of the microscopic dynamics of a system. As such, they provide an alternative to the computation of partition functions: Even though they contain the complete physical information of a system in thermodynamic equilibrium and allow the derivation of all state variables and their relationships, their analytical computation is only technically possible for reasonably simple systems, for instance for ideal gases, but fails for systems with interactions, the Ising-model in three dimensions being a famous example. Sampling  with the Metropolis-Hastings algorithm generates microstates with a relative occurrence given by the Boltzmann-factor $p\propto\exp(-\beta\epsilon)$ with $\beta$ being the inverse temperature and $\epsilon$ the energy difference of the states, effectively sampling from the canonical ensemble.

In parallel, sampling from a posterior distribution with a MCMC-method has become a central tool in cosmology \citep[starting from][]{lewis_cosmological_2002}. There, MCMC-sampling is used to explore the shape of likelihoods as it offers tremendous computational advantages over simple grid-based evaluations, in particular in highly dimensional parameter spaces. Technically, the Metropolis-Hastings algorithm generates parameter tuples $\theta^\mu$ as samples which occur at a probability which is proportional to the posterior distribution $p(\theta|y)$ given the experimental data $y^i$. For this purpose, one requires the logarithmic likelihood, or in the case of a Gaussian error model, the $\chi^2(y|\theta)$-functional, which itself combines the information about the data prediction for a given set of model parameters $\theta^\mu$, as well as the magnitude of the experimental error and the statistical independence of the data in the form of a data covariance. 

A prior distribution $\pi(\theta)\propto\exp(-\phi(\theta))$, reflecting the state of knowledge on the parameters $\theta$ before the measurement, can be incorporated by replacing $\chi^2(y|\theta)/2$ with $\chi^2(y|\theta)/2+\phi(\theta)$. The exponential relationship $\likeli\propto\exp(-\chi^2(y|\theta)/2)$ between $\chi^2(y|\theta)$ and the likelihood $\likeli$ is completely analogous to the probability $p$ for a microstate with energy $\Delta V$ to occur in statistical physics, where a thermal bath is able to provide this energy at the Boltzmann probability $p = \exp(-\beta \Delta V)$. The fact that in statistical inference the Metropolis-Hastings algorithm samples microstates from the partition function can therefore be illustrated with physical analogy: Effectively, $V(\theta) = \chi^2(y|\theta)/2+\phi(\theta)$ plays the role of a potential, on which a particle with coordinates $\theta^\mu$ performs a thermal random walk. Then, the probability of finding the particle at a certain location corresponds to the posterior distribution $p(\theta|y)\propto\exp(-\chi^2(y|\theta)/2-\phi(\theta))$ and is numerically evaluated as the sample density in $\theta^\mu$.

As cosmological likelihoods have in general non-Gaussian shapes and are joint distributions in a large number of parameters with possibly strong degeneracies between them, analytical methods are difficult to employ beyond the second order. As soon as a Gaussian approximation to the likelihood is permissible, though, the Fisher-formalism can play its strength \citep{tegmark_karhunen-loeve_1997}, and has found applications throughout cosmology \citep[for instance,][]{bassett_fisher4cast_2009,bassett_fisher_2011,coe_fisher_2009,elsner_fast_2012, refregier_icosmo:_2011, amara_figures_2011, raveri_cosmicfish_2016}. Extensions to the Fisher-matrix formalism including higher-order derivatives for dealing with non-Gaussian likelihoods have been constructed that are able to capture non-Gaussian shapes of distributions \citep{wolz_validity_2012, giesel_information_2021, schafer_describing_2016, sellentin_breaking_2014} and that can treat systematic errors in addition to statistical errors \citep{amara_systematic_2007, taburet_biases_2009, kitching_path_2011, schafer_weak_2012, kitching_cosmological_2009, loverde_magnification-temperature_2007}.

Motivated by the paper by \citet{jaynes_information_1957}, who clarified that thermodynamics is in reality a theory of information, we aim to continue to use ideas from thermodynamics in statistical inference and to establish analytical methods akin to the partition functions of statistical physics in order to understand the relation between the physical model and the shape of the likelihood. For this purpose, we summarise the construction of canonical partition functions from which a MCMC-algorithm would sample microstates in Sect.~\ref{sect_partitions}. The recovery of the conventional Fisher-formalism in this formalism is demonstrated in Sect.~\ref{sect_nongaussian}, where we show es well the unbiasedness of Gaussian likelihoods and the Cram{\'e}r-Rao limit. The central analytical result of our paper is summarised in Sect.~\ref{sect_gram_charlier}, where we expand the partition function approach beyond the realm of linear models to non-Gaussian posterior distributions and bridge between the DALI-expansion \citep{sellentin_breaking_2014} and the more conventional Gram-Charlier parameterisation. In Sect.~\ref{sect_entropy} we show how information entropies, in particular the Shannon-entropy, are related to the partition functions, recovering a known result from statistical physics along a completely different avenue, and demonstrating the close relationship between Shannon or Kullback-Leibler measures of entropy and the Bayes-theorem. In Sect.~\ref{sect_supernova}, we apply our method to a weakly non-Gaussian likelihood from supernova cosmology \citep{Riess1998, perlmutter_supernovae_2003, goobar_supernova_2011}, specifically constraints on a simple dark energy model \citep[for reviews, see][]{Chung2003, QuintessenceCMB2, mortonson2013dark, AmendolaDarkEnergy, tsujikawa_quintessence:_2013}. With the Union2.1-sample of supernovae, we show that the computation of cumulants of the posterior distribution of the matter density parameter $\Omega_m$ and the dark energy equation of state parameter $w$ directly from Monte-Carlo Markov-chain samples and from differentiation of the partition function are equivalent, and construct an analytic approximation of the posterior distribution with a Gram-Charlier-series. Finally, a discussion in Sect.~\ref{sect_summary} concludes the article.

Throughout the paper, we adopt the summation convention and make the choice to denote parameter tuples $\theta^\mu$ and data tuples $y^i$ as vectors with contravariant indices; Greek indices are reserved for quantities in parameter space and Latin indices for objects in data space. With these conventions, the data covariance $C^{ij} = \bra y^i y^j\ket-\bra y^i\ket\bra y^j\ket$ is a contravariant tensor with $C_{ij}$ as its covariant inverse, $C^{ij}C_{jk} = \delta^i_k$. Similarly, the Fisher-matrix $F_{\alpha\beta}$ is a covariant tensor, allowing to write quadratic forms like $\chi^2 = F_{\alpha\beta}\theta^\alpha\theta^\beta$, and to define an inverse $F^{\alpha\beta}$, $F_{\alpha\beta}F^{\beta\gamma} = \delta_\alpha^\gamma$. 
For an example application in cosmology, we consider supernova-constraints in the frame of a flat $w$CDM-family of FLRW-cosmologies with the matter density $\Omega_m$ and the dark energy equation of state parameter $w$ as parameters. Flatness implies that the density parameter of dark energy is given by $1-\Omega_m$, and it suffices for demonstration purposes to assume a $w$ that is constant in time.

\section{partition functions and Metropolis-Hastings sampling}
\label{sect_partitions}
The Bayes-theorem
\begin{equation}
p(\theta|y) = \frac{\likeli(y|\theta)\pi(\theta)}{p(y)}
\label{eqn_bayes}
\end{equation}
is the foundation of inference as it relates the posterior distribution $p(\theta|y)$ of the model parameter(s) $\theta$ given the data $y$ to the likelihood $\likeli(y|\theta)$ and the prior distribution $\pi(\theta)$. The posterior distribution is properly normalised by the Bayesian evidence $p(y)$,
\begin{equation}
p(y) = \int\dd^n\theta\:\likeli(y|\theta)\pi(\theta).
\end{equation}
For distributions from the exponential family, $\likeli(y|\theta)\propto\exp(-\chi^2(y|\theta)/2)$ with the $\chi^2$-functional and $\pi(\theta)\propto\exp(-\phi(\theta))$ with the logarithmic prior $\phi$ one can reformulate the evidence,
\begin{equation}
p(y) = \int\dd^n\theta\:\exp\left(-\left[\frac{1}{2}\chi^2(y|\theta) + \phi(\theta)\right]\right).
\end{equation}

Enabling Laplace-transforms by introducing sources $J_\alpha$ and scaling the exponent with an inverse temperature $\beta$ leads to an expression reminiscent of a canonical partition function
\begin{equation}
Z[\beta,J_\alpha] = 
\int\dd^n\theta\:\exp\left(-\beta\left[\frac{1}{2}\chi^2(y|\theta) + \phi(\theta)\right]\right)\exp(\beta J_\alpha\theta^\alpha)
\end{equation}
where the (inverse) temperature $\beta$ and the sources $J_\alpha$ play the role of state variables.

Then, the cumulants of the posterior distribution of order $n$ follow from $n$-fold derivation of with respect to the sources $J_\alpha$, i.e. the expectation value is given by
\begin{equation}
\kappa^{\mu} = \bra\theta^\mu\ket = 
\frac{\partial}{\partial J_\mu}\left(\frac{1}{\beta}\ln Z[\beta,J_\alpha]\right)\Bigg|_{J = 0},
\end{equation}
evaluated at $J_\mu = 0$, as well as the covariance by
\begin{equation}
\kappa^{\mu,\nu} = 
\bra\theta^\mu\theta^\nu\ket - \bra\theta^\mu\ket\bra\theta^\nu\ket = 
\frac{\partial^2}{\partial J_\mu \partial J_\nu}\left(\frac{1}{\beta}\ln Z[\beta,J_\alpha]\right)\Bigg|_{J = 0},
\label{eqn_cumulant_2}
\end{equation}
and at higher order corresponding to skewness
\begin{equation}
\kappa^{\mu,\nu,\rho} = 
\bra\theta^\mu\theta^\nu\theta^\rho\ket - \bra\theta^\mu\ket\bra\theta^
\nu\theta^\rho\ket - \bra\theta^\nu\ket\bra\theta^\mu\theta^\rho\ket - \bra\theta^\rho\ket\bra\theta^\mu\theta^\nu\ket + 2\bra\theta^\mu\ket\bra\theta^\nu\ket\bra\theta^\rho\ket =
\frac{\partial^3}{\partial J_\mu\partial J_\nu\partial J_\rho}\left(\frac{1}{\beta}\ln Z[\beta,J_\alpha]\right)\Bigg|_{J = 0},
\label{eqn_cumulant_3}
\end{equation}
and a non-Gaussian kurtosis
\begin{equation}
\kappa^{\mu,\nu,\rho,\sigma} = 
\bra\theta^\mu\theta^\nu\theta^\rho\theta^\sigma\ket - \bra\theta^\mu\theta^\nu\ket\bra\theta^\rho\theta^\sigma\ket - \bra\theta^\mu\theta^\rho\ket\bra\theta^\nu\theta^\sigma\ket - \bra\theta^\mu\theta^\sigma\ket\bra\theta^\nu\theta^\rho\ket = 
\frac{\partial^4}{\partial J_\mu\partial J_\nu\partial J_\rho\partial J_\sigma}\left(\frac{1}{\beta}\ln Z[\beta,J_\alpha]\right)\Bigg|_{J = 0},
\end{equation}
all taken at $J_\mu = 0$ after differentiation; the dependence on temperature $\beta$ is lost due to the differentiation operating on $\ln Z[\beta,J_\alpha]/\beta$. It is remarkable that differentiation of the partition function yields the expectation values of the parameters $\bra\theta^\mu\ket$ and not the best-fit values.

Commonly, one uses Monte-Carlo Markov-chain methods such as the Metropolis-Hastings algorithm to sample microstates which correspond to parameter tuples $\theta^\mu$ from the partition function $Z$. From these samples one can compute estimates for the cumulants, or equivalently, of the moments, as the two are related by Faa di Bruno's formula \citep{johnson_curious_2002}. Here, we advocate instead that the cumulants up to the order of non-Gaussianity that one decides to truncate the expansion, are given by possibly numerical differentiation of the logarithmic partition function, and that this approach yields an analytic approximation to the posterior distribution $p(\theta|y)$.

\section{(non)linear models and (non)Gaussian likelihoods}
\label{sect_nongaussian}

\subsection{Linear models}
Linear models together with a Gaussian error process on the data have particularly convenient properties in data analysis. The $\chi^2$-functional for data $y^i$ with an inverse data covariance $C_{ij}$, resulting from $C^{ij} = \bra y^iy^j\ket - \bra y^i\ket\bra y^j\ket$, for a model that is linear in the parameters, $y_{\text{model}}^i = A\indices{^i_\alpha}\theta^\alpha$ is given by a quadratic form
\begin{equation}
\chi^2 = 
\left(y^i - A\indices{^i_\alpha}\theta^\alpha\right) C_{ij} \left(y^j - A\indices{^j_\beta}\theta^\beta\right) = 
\underbrace{y^iC_{ij}y^j}_{c} - 2\underbrace{y^jC_{ij}A\indices{^i_\alpha}}_{Q_\alpha}\theta^\alpha + \underbrace{A\indices{^i_\alpha}C_{ij}A\indices{^j_\beta}}_{F_{\alpha\beta}}\theta^\alpha\theta^\beta = 
c - 2Q_\alpha\theta^\alpha + F_{\alpha\beta}\theta^\alpha\theta^\beta.
\end{equation}
The quadratic term $A\indices{^i_\alpha}C_{ij}A\indices{^j_\beta}$ is identified as the Fisher-matrix $F_{\alpha\beta}$, with the Jacobian $A\indices{^i_\alpha} = \partial y^i/\partial\theta^\alpha$ taking care of the change of basis between parameter space and data space. Consequently, likelihoods of linear models are Gaussian in the data as well as in the parameters,
\begin{equation}
\likeli(y|\theta) \propto \exp\left(-\frac{1}{2}\chi^2(y|\theta)\right).
\end{equation}
and unless the prior $\pi(\theta)$ deviates from being constant or Gaussian itself, the posterior distribution $p(\theta|y)$ will be Gaussian, too.

The best-fit parameter tuple $\bar{\theta}^\mu$ is readily computed by minimisation of $\chi^2$ as a function of the parameters $\theta^\mu$,
\begin{equation}
\frac{\partial}{\partial\theta^\mu} \chi^2 = 
-2Q_\alpha\frac{\partial\theta^\alpha}{\partial\theta^\mu} + F_{\alpha\beta}\frac{\partial}{\partial\theta^\mu}\theta^\alpha\theta^\beta = 
-2Q_\alpha\delta^\alpha_\mu + F_{\alpha\beta} \left(\delta^\alpha_\mu\theta^\beta + \theta^\alpha\delta^\beta_\mu\right) = 
-2Q_\mu + F_{\mu\beta}\theta^\beta + F_{\alpha\mu}\theta^\alpha = 0
\quad\rightarrow\quad
\bar{\theta}^\mu = F^{\mu\alpha}Q_\alpha = A\indices{^\mu_i} y^i
\label{eqn_quadratic_estimate}
\end{equation}
recovering estimation with the inverse Jacobian $A\indices{^\alpha_i} = \partial\theta^\alpha/\partial y^i$: In fact, the estimate is unbiased because the true model $\hat{\theta}^\mu$ would generate the data $ y^i = A\indices{^i_\mu}\hat{\theta}^\mu$, such that $\bar{\theta}^\mu = A\indices{^\mu_i} y^i = A\indices{^\mu_i}A\indices{^i_\nu}\hat{\theta}^\nu = \delta^\mu_\nu\hat{\theta}^\nu = \hat{\theta}^\mu$. 

Constructing the partition function for such a linear model amounts to
\begin{equation}
Z[\beta,J_\alpha] = 
\int\dd^n\theta\:\exp\left(-\frac{\beta}{2}F_{\alpha\beta}\theta^\alpha\theta^\beta + \beta Q_\alpha \theta^\alpha\right)\exp\left(\beta J_\alpha\theta^\alpha\right).
\end{equation}
where the Gaussian integrals can be carried out to yield
\begin{equation}
Z[\beta,J_\alpha] = 
\sqrt{\left(\frac{2\pi}{\beta}\right)^n\frac{1}{\mathrm{det}(F)}}
\exp\left(\frac{\beta}{2}F^{\alpha\beta}(J_\alpha+Q_\alpha)(J_\beta + Q_\beta\right)
\label{eqn_linear_model}
\end{equation}
with the inverse Fisher-matrix $F^{\alpha\beta}$. Here, we have dropped the irrelevant normalisation $\exp(-c/2)$ of the likelihood and disregarded any prior $\pi(\theta)$ for simplicity. 

The expectation value of the posterior distribution, or equivalently, the first cumulant $\kappa^\mu$ follows directly from differentiation of $\ln Z[\beta,J_\alpha]/\beta$, evaluated at $J_\mu = 0$, keeping in mind that the Fisher-matrix and its inverse are symmetric,
\begin{equation}
\kappa^\mu = 
\frac{\partial}{\partial J_\mu}\left(\frac{1}{\beta}\ln Z[\beta,J_\alpha]\right)\Bigg|_{J = 0} = 
\frac{F^{\alpha\beta}}{2}\left(\frac{\partial J_\alpha}{\partial J_\mu}(J_\beta + Q_\beta) + (J_\alpha + Q_\alpha)\frac{\partial J_\beta}{\partial J_\mu}\right)\Bigg|_{J = 0} =
\frac{F^{\alpha\beta}}{2}\left(\delta^\mu_\alpha(J_\beta + Q_\beta) + (J_\alpha + Q_\alpha)\delta^\mu_\beta\right)\Bigg|_{J = 0} =
F^{\mu\alpha}Q_\alpha
\end{equation}
where one recovers the result from the direct calculation in eqn.~(\ref{eqn_quadratic_estimate}): For a symmetric distribution, most likely value and expectation value coincide with the true parameter value: $\kappa^\mu = \bra\theta^\mu\ket = \hat{\theta}^\mu$, and incidentally $\hat{\theta^\mu} = \bar{\theta}^\mu$, too, as a reflection of the Gau\ss-Markov theorem in this formalism.

The second cumulant $\kappa^{\mu,\nu}$, corresponding to the parameter covariance, is computed as
\begin{equation}
\kappa^{\mu,\nu} = 
\frac{\partial^2}{\partial J_\mu\partial J_\nu}\left(\frac{1}{\beta}\ln Z[\beta,J_\alpha]\right)\Bigg|_{J = 0} = 
\frac{F^{\alpha\beta}}{2}\left(\delta^\mu_\alpha\delta^\nu_\beta + \delta^\nu_\alpha\delta^\mu_\beta\right) \Bigg|_{J = 0} = F^{\mu\nu}
\end{equation}
Consequently, as any higher-order cumulants are zero, the posterior distribution $p(\theta|y)$ is necessarily Gaussian, with the Fisher-matrix as the inverse covariance and is centered on the expectation value:
\begin{equation}
p(\theta|y) = 
\sqrt{\frac{\mathrm{det}(F)}{(2\pi)^n}}\exp\left(-\frac{1}{2}F_{\mu\nu}(\theta^\mu-\kappa^\mu)(\theta^\nu-
\kappa^\nu)\right),
\end{equation}
because $F_{\mu\alpha}F^{\alpha\nu} = F_{\mu\alpha}\kappa^{\alpha,\nu} = \delta_\mu^\nu$.

\subsection{Cram{\'e}r-Rao inequality}
Generally, the (co)variance of posterior distributions $p(\theta|y)$ is bounded from below by the inverse $1/F$ of the Fisher-information $F$, as stated by the Cram{\'e}r-Rao inequality,
\begin{equation}
\kappa^{(2)} \geq \frac{1}{F}
\quad\text{with}\quad
F = \left\langle\frac{\partial^2}{\partial\theta^2}\frac{\chi^2(y|\theta)}{2}\right\rangle,
\end{equation}
averaged over the parameters $\theta$, where the logarithmic likelihood $\ln\likeli$ corresponds for a Gaussian error process to $-\chi^2/2$. The partition-function formalism provides a way to derive this relationship because any cumulant of the posterior distribution is given by differentiation of $\ln Z[\beta,J]/\beta$ with respect to the source terms $J$ and because the partition function $Z$ reflects the specific functional dependence of $\chi^2(y|\theta)$ on the parameters $\theta$ themselves. Therefore, it is sensible to ask whether there is a particular $\chi^2(y|\theta)$ that minimises the variance for a fixed value of the Fisher information. Essentially, this amounts to carrying out a functional minimisation of $\kappa^{(2)}$ with respect to $\chi^2(y|\theta)/2$:
\begin{equation}
\delta\kappa^{(2)} = 
\delta\left[\bra\theta^2\ket-\bra\theta\ket^2\right] = 
\delta\frac{\partial^2}{\partial J^2}\ln Z[J]\Bigg|_{J = 0} =
\frac{\partial^2}{\partial J^2}\delta\ln Z[J]\Bigg|_{J = 0} = 
\frac{\partial^2}{\partial J^2}\frac{\delta Z[J]}{Z[J]}\Bigg|_{J = 0} = 0
\end{equation}
where we focus for now on unit temperature $\beta = 1$ and on the univariate case, for simplicity, as well as a constant prior distribution $\pi(\theta)$.

Fixing the value of the Fisher information to $F$, the expectation value of the distribution to $\bra\theta\ket$ and the normalisation to one can be done by incorporating three Lagrange-multipliers $\alpha$, $\beta$ and $\gamma$,
\begin{equation}
\delta\kappa^{(2)}\rightarrow
\delta\kappa^{(2)} + 
\alpha\:\delta\left[\left\bra\frac{\partial^2}{\partial\theta^2}\frac{\chi^2(y|\theta)}{2}\right\ket - F\right] +
\beta\:\delta\left[\left\bra\theta\right\ket - \mu\right] + 
\gamma\:\delta\left[\int\dd\theta\:\exp\left(-\frac{\chi^2(y|\theta)}{2}\right) - 1\right].
\end{equation}
Collecting all terms leads to the variation
\begin{equation}
\int\dd\theta\:\exp\left(-\frac{\chi^2(y|\theta)}{2}\right)\left[
\theta^2 + 2\bra\theta\ket\:\theta + \alpha\left(2\frac{\partial^2}{\partial\theta^2}\frac{\chi^2(y|\theta)}{2} - \left(\frac{\partial}{\partial\theta}\frac{\chi^2(y|\theta)}{2}\right)^2\right) + \beta \theta + \gamma
\right]\:\delta\left(-\frac{\chi^2}{2}\right) = 0
\end{equation}
using the relation $\delta(\bra\theta\ket^2) = 2\bra\theta\ket\:\delta\bra\theta\ket$ and the fact that every term is proportional to $1/Z$, such that the term in the square brackets needs to vanish. Abbreviating the negative logarithmic likelihood as $q = \chi^2(y|\theta)/2$ and using primes to denote the derivative with respect to $\theta$ yields the differential equation
\begin{equation}
\theta^2 - 2\bra\theta\ket\:\theta + \alpha(2q^{\prime\prime} - {q^\prime}^2) + \beta\theta + \gamma = 0.
\label{eqn_cr_deq}
\end{equation}
As a polynomial, the solution $q$ can only be consistent if its second derivative and the square of its first derivative do no generate powers of $\theta$ higher than 2. Therefore, a sensible ansatz would be
\begin{equation}
q = \frac{\chi^2(y|\theta)}{2} = \frac{1}{2}F(\theta-\mu)^2
\end{equation}
which is readily shown to solve the variational problem with $\alpha = 1/F^2$, $\beta = 0$ and the (irrelevant) normalisation $\gamma = \mu - 2/F$. Substituting back into the expression for $\kappa^{(2)}$
leads directly to the value $1/F$, in fulfilment of the Cram{\'e}r-Rao inequality.

The multivariate case is slightly more complicated and requires $n$ individual Lagrange-multipliers $\beta^\nu$ for enforcing the means $\bra\theta^\mu\ket$ of the distribution, replacing eqn.~(\ref{eqn_cr_deq}) by
\begin{equation}
\theta^\mu\theta^\nu - (\bra\theta^\mu\ket\theta^\nu + \theta^\mu\bra\theta^\nu\ket) + 
\alpha n\left(2\frac{\partial^2q}{\partial\theta^\mu\partial\theta^\nu} - \frac{\partial q}{\partial\theta^\mu}\frac{\partial q}{\partial\theta^\nu}\right) + 
\beta^\nu\theta^\mu + \gamma = 0,
\quad\text{with the definition}\quad
F_{\mu\nu} = n\left\bra\frac{\partial^2}{\partial\theta^\mu\partial\theta^\nu}\frac{\chi^2(y|\theta)}{2}\right\ket.
\end{equation}
Then, the same arguments as in the univariate case lead to a quadratic form
\begin{equation}
q = \frac{\chi^2(y|\theta)}{2} = 
\frac{1}{2} F_{\mu\nu}(\theta^\mu - \bra\theta^\mu\ket)(\theta^\nu - \bra\theta^\nu\ket).
\end{equation}

In summary, the minimum variance for a fixed Fisher-information is realised by a quadratic $\chi^2(y|\theta)$-functional, or equivalently, a Gaussian posterior distribution $p(\theta|y)$. Commonly, $\chi^2$ is a convex function of the parameters $\theta^\mu$, otherwise there can not be a global minimum for uniquely defining the best fitting parameters $\bar{\theta}^\mu$. Consequently, the functional to be minimised is convex, too: $\exp(-\chi^2/2)$ is convex if $\chi^2$ is convex. Multiplying $\exp(-\chi^2)$ with strictly positive functions $q^{\prime\prime}$ or $\theta^2$ does not change convexity, which implies that the functional variation has yielded a minimum for the variance for a given Fisher information $F$.

\subsection{Nonlinear models}
Nonlinear models are notably more complicated: Neither is the $\chi^2$-functional quadratic in the parameters nor is the likelihood $\likeli$ of Gaussian shape, simply because the function $y^i(\theta^\alpha)$ can not be written as $y^i = A\indices{^i_\alpha}\theta^\alpha$ with a constant $A\indices{^i_\alpha}$, or equivalently, because $\partial y^i/\partial\theta^\alpha$ becomes a function of $\theta^\alpha$ itself. But in the language of partition functions, this means that they are neither of Gaussian form, nor would the differentiations truncate after second order: The posterior distribution becomes genuinely non-Gaussian. But the cumulants of order $n$ with $n\geq3$ remain computable, at least numerically, from the partition function. In this sense, the partition function formalism provides an approximation for non-Gaussian posterior distributions at a given order. At least for weak non-Gaussianities, which would be the case if a linearisation of the physical model over the allowed parameter space is applicable, the partition function can be decomposed into a Gaussian part and a non-Gaussian correction, written in terms of Hermite-polynomials, linking to the Gram-Charlier-expansion for weakly non-Gaussian distributions. From this moment on, we always assume a rescaling of the model such that the fiducial parameter values are zero, $\bra\theta^\mu\ket = \kappa^\mu = 0$, simplifying the notation tremendously.

\section{weakly non-Gaussian likelihoods}\label{sect_gram_charlier}

\subsection{Higher cumulants}
	When introducing a weak non-Gaussianity, e.g. through introducing a model where the parameters are not quite linearly linked to the data the partition function factorizes into a Gaussian and a non-Gaussian part. The non-Gaussianity is introduced in the $\chi^2$-functional as 
	\begin{align}
\frac{\chi^2}{2} = \frac{1}{2}F_{\alpha \beta} \theta^\alpha \theta^{\beta} - \sum_{k=3}^{N}  \frac{1}{k!}C_{\mu_1 \dots \mu_k} \theta^{\mu_1}\dots \theta^{\mu_k},
	\end{align}
i.e. with a Taylor-expansion of $\chi^2$ beyond quadratic order, as it would naturally arise in nonlinear models. The coefficients in the Taylor expansion are assumed to be small compared to the entries of the covariance matrix $F_{\alpha \beta}$, and the minus-sign of the nonlinear term $C_{\mu_1 \dots \mu_k}$ is chosen out of convenience. This allows to separate the Gaussian from the non-Gaussian part in the partition function.
	\begin{align}
		Z[\beta, J_\alpha] &= \int\dd^n\theta \exp\left( -\frac{\beta}{2} F_{\alpha \beta} \theta^\alpha \theta^\beta + \beta\sum_{k=3}^{N}  \frac{1}{k!}C_{\mu_1 \dots \mu_k} \theta^{\mu_1}\dots \theta^{\mu_k} + \beta J_\alpha \theta^\alpha\right)\approx  \int\dd^n\theta \exp\left( -\frac{\beta}{2} F_{\alpha \beta} \theta^\alpha \theta^\beta + \beta J_\alpha \theta^\alpha\right) \left(1 + \beta\sum_{k=3}^{N}  \frac{1}{k!}C_{\mu_1 \dots \mu_k} \theta^{\mu_1}\dots \theta^{\mu_k}\right)
		\label{factorization_approximation}
	\end{align}
	
In a next step the computation of the moments is replaced with a differentiation with respect to the sources $J_\alpha$. The partition function can then be expressed as 
	\begin{align}
		Z[\beta, J_\alpha] &= \left(1 + \beta\sum_{k=3}^{N}  \frac{1}{k!}C_{\mu_1 \dots \mu_k} \frac{\partial}{\partial J_{\mu_1}}\dots \frac{\partial}{\partial J_{\mu_k}}\right)\int\dd^n\theta \exp\left( -\frac{\beta}{2} F_{\alpha \beta} \theta^\alpha \theta^\beta + \beta J_\alpha \theta^\alpha\right)\\
		&=\sqrt{\left(\frac{2\pi}{\beta}\right)^n\frac{1}{\mathrm{det}(F)}} \left(1 + \beta\sum_{k=3}^{N}  \frac{1}{k!}C_{\mu_1 \dots \mu_k} \frac{\partial}{\partial J_{\mu_1}}\dots \frac{\partial}{\partial J_{\mu_k}}\right) \exp\left(\frac{\beta}{2}F^{\alpha\beta}J_\alpha J_\beta\right)\\
		&=\sqrt{\left(\frac{2\pi}{\beta}\right)^n\frac{1}{\mathrm{det}(F)}}\exp\left(\frac{\beta}{2}F^{\alpha\beta}J_\alpha J_\beta\right) \left(1 + \beta\sum_{k=3}^{N}  \frac{1}{k!}C_{\mu_1 \dots \mu_k} \exp\left(-\frac{\beta}{2}F^{\alpha\beta}J_\alpha J_\beta\right)\frac{\partial}{\partial J_{\mu_1}}\dots \frac{\partial}{\partial J_{\mu_k}}\exp\left(\frac{\beta}{2}F^{\alpha\beta}J_\alpha J_\beta\right)\right) 
	\end{align}
	The first factor in this equation is equivalent to the partition function for a linear model as given in eqn. (\ref{eqn_linear_model}). Here the mean values are set to zero. The second factor in the equation is the contribution of non-Gaussianities to the partition function. With the definitions
	\begin{equation}
		Z_G[\beta,J_\alpha] = \sqrt{\left(\frac{2\pi}{\beta}\right)^n\frac{1}{\mathrm{det}(F)}}\exp\left(\frac{\beta}{2}F^{\alpha\beta}J_\alpha J_\beta\right) 
	\end{equation}
	as well as
	\begin{equation}
		Z_{NG}[\beta,J_\alpha] = \left(1 + \beta\sum_{k=3}^{N}  \frac{1}{k!}C_{\mu_1 \dots \mu_k} \exp\left(-\frac{\beta}{2}F^{\alpha\beta}J_\alpha J_\beta\right)\frac{\partial}{\partial J_{\mu_1}}\dots \frac{\partial}{\partial J_{\mu_k}}\exp\left(\frac{\beta}{2}F^{\alpha\beta}J_\alpha J_\beta\right)\right)
	\end{equation} 
	a factorization of the partition function into a contribution due to its Gaussian part and the influence of the non-Gaussianities can be observed,
	\begin{align}
		Z[\beta,J_\alpha]) = Z_G[\beta,J_\alpha]Z_{NG}[\beta,J_\alpha].
	\end{align}
    Using the fact that the inverse Fisher matrix can be Cholesky decomposed as $F^{\alpha \beta} = L^{\gamma \beta} L\indices{_\gamma^\alpha}$ the non-Gaussian part can be expressed in terms of multivariate Hermite polynomials of the form 
	\begin{align}
		H^{(\nu_1 \dots \nu_\ell)} (\mathbf{J}) &= \exp \left(\frac{1}{2}J_\alpha \delta^{\alpha \beta}
		J_\beta \right)(-1)^\ell \frac{\partial^\ell}{\partial J_{\nu_1}\dots \partial J_{\nu_\ell}}\exp \left(-\frac{1}{2}J_\alpha \delta^{\alpha \beta}
		J_\beta \right)
		\label{eqn:hermite_factorizing}
	\end{align} 
    as
	\begin{align}
		Z_{NG}[\beta,J_\alpha] = \left(1 + \beta\sum_{k=3}^{N}  \frac{(-\ci^k)}{k!}C_{\mu_1 \dots \mu_k}L_{\nu_1}^{\mu_1}\dots L_{\nu_k}^{\mu_k} H^{(\nu_1 \dots \nu_k)}(\ci\mathbf{LJ})\right).
	\end{align}
    When computing cumulants the logarithm of these quantities is considered and the contributions of the Gaussian and non-Gaussian part can be expressed as a sum. Up to first order in the non-Gaussianities the expression is
	
	\begin{align}
		\ln Z[\beta,J_\alpha] &= \ln Z_G[\beta,J_\alpha] + \ln Z_{NG}[\beta,J_\alpha] 
		\approx \frac{\beta}{2}F^{\alpha\beta}J_\alpha J_\beta + \beta\sum_{k=3}^{N}  \frac{(-\ci^k)}{k!}C_{\mu_1 \dots \mu_k}L_{\nu_1}^{\mu_1}\dots L_{\nu_k}^{\mu_k} H^{(\nu_1 \dots \nu_k)}(\ci\mathbf{LJ}) + \mathrm{const}.
	\end{align}
    Note that in eqn.~(\ref{factorization_approximation}) the approximation can be performed to higher order in the non-Gaussianities for the cost of including higher order Hermite polynomials in the result. The factorization itself can still be performed. 
	
	For non-vanishing expectation values such that 
	\begin{align}
		\frac{\chi^2}{2} = \frac{1}{2}F_{\alpha \beta} \theta^\alpha \theta^{\beta} + \mu_\alpha \theta^{\alpha}  - \sum_{k=3}^{N}  \frac{1}{k!}C_{\mu_1 \dots \mu_k} \theta^{\mu_1}\dots \theta^{\mu_k}
	\end{align}
	the Gaussian term is modified by a term linear in the sources
	\begin{align}
		\ln Z_G[\beta,J_\alpha] = 
		\frac{\beta}{2} F^{\alpha \beta} J_\alpha J_\beta - \beta \mu^{\alpha} J_\alpha,
	\end{align}
	and consequently, the non-Gaussian term in the partition function remains unchanged. The higher order cumulants of the posterior can be computed as 
\begin{align}
\kappa^{\mu_1,\dots ,\mu_n} = 
\frac{\partial^n}{\partial J_{\mu_1} \dots \partial J_{\mu_n}}\left(\frac{1}{\beta}\ln Z[\beta,J_\alpha]\right)\Bigg|_{J = 0} = 
\frac{\partial^n}{\partial J_{\mu_1} \dots \partial J_{\mu_n}} \left(\frac{1}{2} F^{\alpha \beta} J_\alpha J_\beta - \sum_{k=3}^{N}  \frac{(-\ci^k)}{k!}C_{\mu_1 \dots \mu_k}L_{\nu_1}^{\mu_1}\dots L_{\nu_k}^{\mu_k} H^{(\nu_1 \dots \nu_k)}(\ci\mathbf{LJ})\right)\Bigg|_{J = 0}.
\label{eqn:cumulants_nG}
\end{align}

The first and second cumulants  already contain contributions from the odd and even Hermite polynomials in the non-Gaussian term respectively. The higher cumulants are completely determined by the non-Gaussian term. The question whether non-Gaussianities are genuine or just an artefact of an unfortunate choice of random variables and could therefore in principle removed by a suitable coordinate transform can be traced to the existence of curvature on the manifold whose metric is given by the Fisher-matrix $F_{\mu\nu}$ \citep[see the foundational work by][]{amari_information_2016}. Applications in cosmology are discussed in \citet{giesel_information_2021}, and variations of the Fisher-matrix over the parameter manifold and the resulting non-Gaussianities in \citet{schafer_describing_2016} and \citet{reischke_variations_2016}.

	\subsection{Gram-Charlier series}
	The cumulants of a weakly non-Gaussian posterior allow to reconstruct it using the multivariate Gram-Charlier series. It can be written as \citep{multivariate_gc, 1995ApJ...442...39J, giesel_information_2021}
	\begin{align}
p(\theta|y) = 
\frac{1}{(2\pi)^{\frac{n}{2}}\sqrt{\det(\kappa^{(2)})}} \exp \left(-\frac{1}{2} \kappa_{\alpha, \beta}(\theta^\alpha - \kappa^\alpha)(\theta^\beta- \kappa^\beta)\right)\left( 1 + \sum_{\ell=3}^{\infty} \frac{\kappa^{\alpha_1, \dots ,\alpha_\ell}}{\ell !} L_{\alpha_1}^{\beta_1} \dots L_{\alpha_\ell}^{\beta_\ell} H_{(\beta_1 \dots \beta_\ell)}(\mathbf{L(\theta-\mu)})\right),
		\label{eqn:gram_chal}
	\end{align}
where the mean value of the distribution is chosen as the first cumulant of the posterior while the covariance matrix is the inverse of the second cumulant.
	The Hermite polynomials in this expression again follow the definition in eqn.~(\ref{eqn:hermite_factorizing}). Due to the covariance matrix chosen as unity in this definition the multidimensional Hermite polynomials factorize into one dimensional ones. The variable transformation necessary to achieve this form can be seen in the argument of the Hermite polynomials. This accounts for the Cholesky decomposed second cumulants in their prefactors.
	
\begin{figure}
		\centering
		\includegraphics[width =0.3 \textwidth]{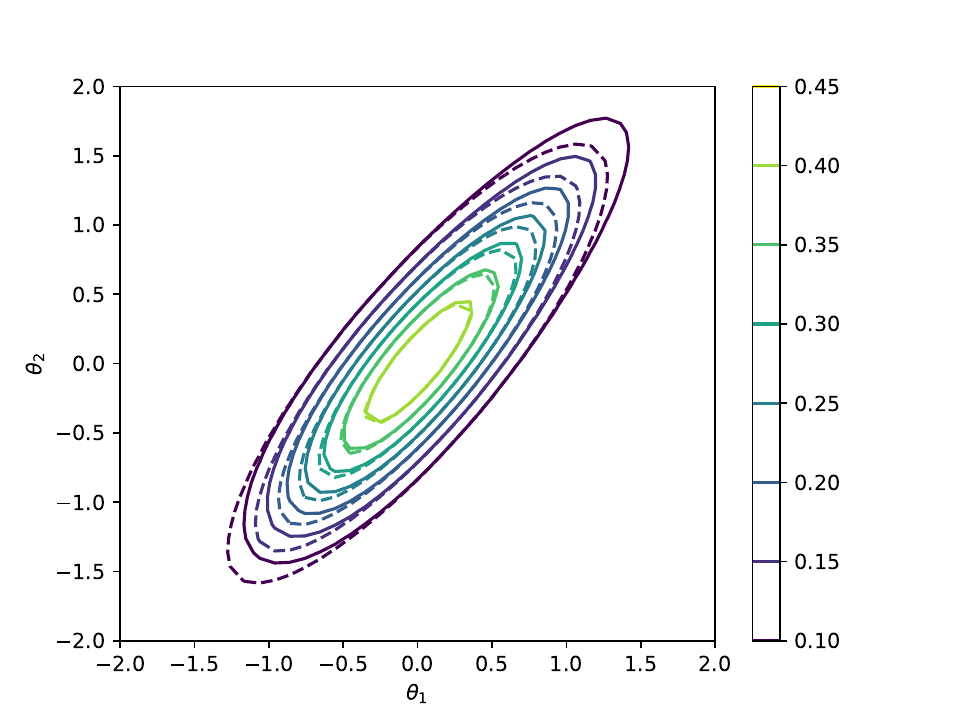}
		\includegraphics[width =0.3 \textwidth]{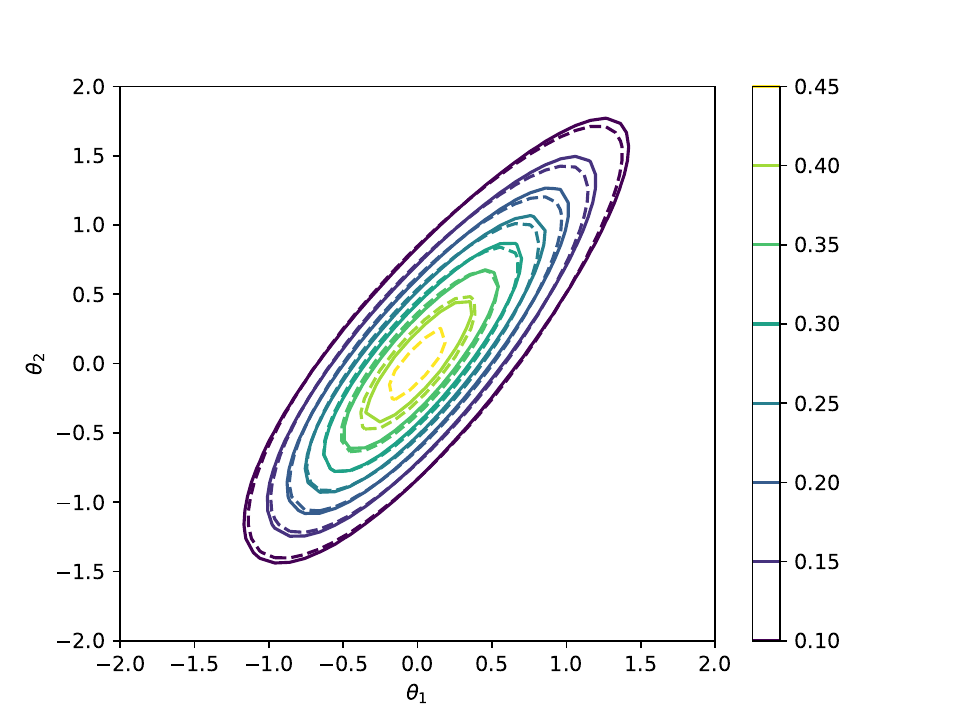}
		\includegraphics[width =0.3 \textwidth]{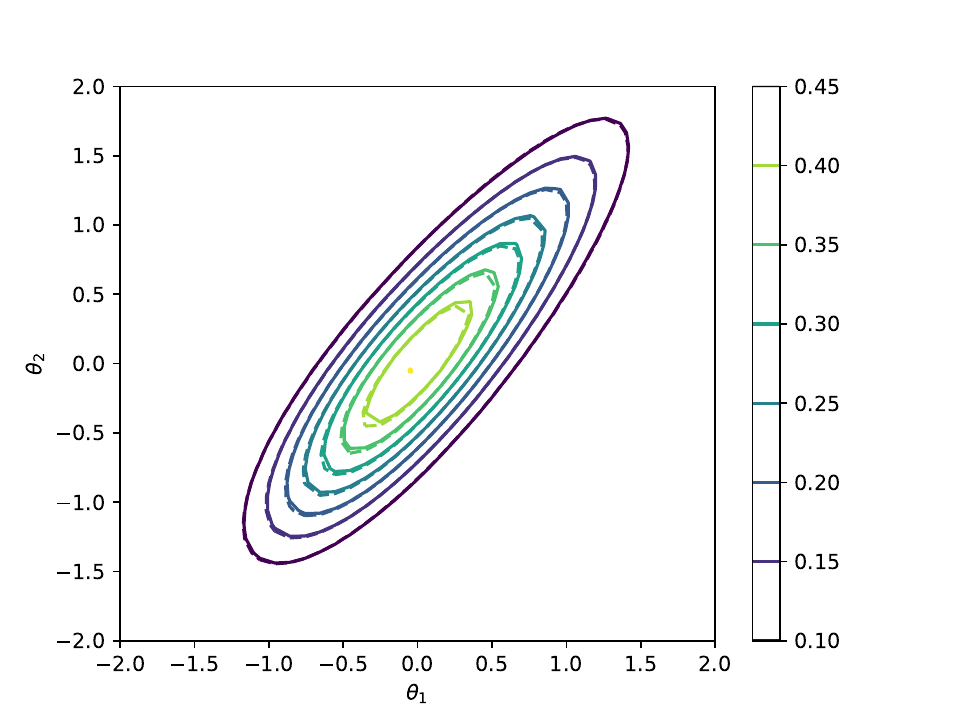}
		\caption{Reconstruction of non-Gaussianity of second order using the Gram-Charlier series. The left plot shows the posterior constructed according to eqn.~(\ref{eqn:mock_example}) and a Gaussian (dashed) with the same width in comparison. The middle (dashed) is the Gram-Charlier series to first order in the non-Gaussianities. The dashed lines in the right depict the Gram-Charlier series to second order in the non-Gaussianities}\label{fig:nonG_3rd}
	\end{figure}

    In the following the Gram-Charlier series is used to reconstruct a probability distribution of the form
    \begin{align}
        p(\theta|y) = 
        \frac{1}{(2\pi)^{\frac{n}{2}}\sqrt{\det(\Sigma)}}\exp \left(-\frac{1}{2} \Sigma_{\alpha \beta} \theta^\alpha \theta^\beta \right)\left(1 + \frac{1}{3!}C_{\alpha \beta \gamma}\theta^\alpha \theta^\beta \theta^\gamma \right).
        \label{eqn:mock_example}
	\end{align}
	Note that this is a first order approximation in the coefficients $C_{\alpha \beta \gamma}$ to a posterior with a third order non-Gaussianity. For small coefficients it is possible to compute the cumulants of this posterior according to eqn.~(\ref{eqn:cumulants_nG}) they can then be inserted into expression eqn.~(\ref{eqn:gram_chal}). The results of that reconstruction are depicted in Fig.~\ref{fig:nonG_3rd} for a toy model of non-Gaussianity.
	
	As demonstrated in the above example it is possible to approximately construct cumulants for a given expansion coefficient in the DALI expansion. However, it is worth noting that even the first order in the DALI expansion does not correspond to any specific order in the Gram-Charlier expansion. This can be seen by considering  
	\begin{align}
    	\exp\left(-\frac{1}{2} F_{\alpha \beta} \theta^{\alpha} \theta^\beta + \frac{1}{3!} S_{\alpha \beta \gamma} \theta^{\alpha} \theta^\beta \theta^\gamma \right) = \exp\left(-\frac{1}{2} F_{\alpha \beta} \theta^{\alpha} \theta^\beta\right) \sum_{n=0}^\infty \frac{1}{n!} (S_{\alpha \beta \gamma} \theta^{\alpha} \theta^\beta \theta^\gamma)^n.
    \end{align}
	Terms in this expansion that are of order $3^n$ in the parameters $\theta$ can only be obtained by including an ever increasing order expansion in the Gram-Charlier series. In this sense the Gram-Charlier series seems to be incompatible with a straightforward expansion of the logarithmic likelihood in terms of a polynomial as done in DALI \citep{sellentin_breaking_2014, sellentin_fast_2015}. One clearly needs a large number of terms to be able to describe even moderately non-Gaussian distributions, set aside issues with non-positive definite probability densities, as pointed out in \citet{cramer1999mathematical}.

\section{partitions, Bayesian evidences and entropy measures}
\label{sect_entropy}

\subsection{Entropy measures}
The partition function $Z[\beta,J_\alpha]$ evaluated at unit temperature $\beta = 1$ and for vanishing sources $J_\alpha = 0$ falls back onto the Bayes evidence as the normalising factor for the posterior distribution,
\begin{equation}
Z[\beta,J_\alpha = 0] = 
\int\dd^n\theta\:\exp\left(-\beta\left[\frac{1}{2}\chi^2(y|\theta)+\phi(\theta)\right]\right)
\quad\text{such that}\quad
Z[\beta = 1,J_\alpha = 0] = p(y),
\end{equation}
which is exactly the normalisation of the posterior distribution $p(\theta|y)$, according to eq.~(\ref{eqn_bayes}). It is an interesting thought to derive the Shannon-entropy $S$ of the posterior distribution,
\begin{equation}
S = -\int\dd^n\theta\: p(\theta|y) \ln p(\theta|y)
\end{equation}
from the partition function $Z[\beta]$, where we disregard the dependence on $J_\alpha$ and set the sources preemptively to zero. In fact, motivated by statistical physics one would define an analogous Helmholtz free energy $F(\beta)$ as the thermodynamical potential corresponding to $Z[\beta]$ through
\begin{equation}
F(\beta) = -\frac{1}{\beta}\ln Z[\beta].
\end{equation}
Then, the entropy $S(\beta)$ follows from $F(\beta)$ through differentiation with respect to temperature, i.e. 
\begin{equation}
S(\beta) = 
\beta^2\frac{\partial F(\beta)}{\partial\beta} = 
-\beta^2\frac{\partial}{\partial\beta}\left(\frac{1}{\beta}\ln Z[\beta]\right) =
\ln Z[\beta] - \beta\frac{\partial}{\partial\beta}\ln Z[\beta].
\end{equation}

The two terms are readily evaluated at $\beta = 1$ to be the logarithmic evidence $\ln p(y)$ as well as 
\begin{equation}
\beta\frac{\partial}{\partial\beta}\ln Z[\beta]\Bigg|_{\beta = 1} = 
-\frac{1}{p(y)}\int\dd^n\theta\:\exp\left(-\left[\frac{1}{2}\chi^2(y|\theta) + \phi(\theta)\right]\right)\times\left[\frac{1}{2}\chi^2(y|\theta) + \phi(\theta)\right]
\end{equation}
as the expectation value of $\chi^2(y|\theta)/2+\phi(\theta)$. With the logarithmic Bayes-law for exponential distributions
\begin{equation}
\ln p(\theta|y) = -\left[\frac{1}{2}\chi^2(y|\theta) + \phi(\theta)\right] - \ln p(y)
\end{equation}
one then obtains with the normalisation $\int\dd^n\theta\: p(\theta|y) = 1$
\begin{equation}
\beta\frac{\partial}{\partial\beta}\ln Z[\beta]\Bigg|_{\beta = 1} = \int\dd^n\theta\: p(\theta|y)\ln p(\theta|y) + \ln p(y).
\end{equation}
Collecting all terms then yields the final result, relating the entropy $S$ directly to Shannon's measure of information entropy \citep[for applications in cosmology, see][]{carron_probe_2011, grandis_information_2016, Pinho:2020uzv, nicola_consistency_2018}
\begin{equation}
S(\beta = 1) = 
-\beta^2\frac{\partial}{\partial\beta}\left(\frac{1}{\beta}\ln Z[\beta]\right)\Bigg|_{\beta = 1} = 
-\int\dd^n\theta\:p(\theta|y)\ln p(\theta|y)
\end{equation}
Interestingly, this is another pathway of showing the compatibility of Shannon's entropy over the wider class of R{\'e}nyi-entropies with respect to Bayes' law \citep{van_erven_renyi_2014, baez_bayesian_2014}.

In summary, there is the remarkable result that at unit temperature, the logarithmic partition function corresponds to the Bayes evidence, and that the derivative of the partition function yields the information entropy of the posterior distribution, illustrating how deeply the Bayes-law and information theory are integrated. At this point, we would like to point out that the computation of Bayesian evidences as absolute numbers for the purpose of model selection \citep{jenkins_power_2011, liddle_present_2006, kerscher_model_2019, handley_quantifying_2019, trotta_bayesian_2017, knuth_bayesian_2015, trotta_bayes_2008, trotta_applications_2007, bellini_constraints_2016, joudaki_cfhtlens_2017, grassi_test_2014}, and their comparison on e.g. the Jeffrey's scale requires the inclusion of the respective normalisation of $\likeli(y|\theta)$ and $\pi(\theta)$. With this result we suggest to compute information entropies through derivatives with respect to temperature of $\ln Z[\beta]/\beta$, which effectively governs the acceptance rate for a given $\Delta\chi^2$ in MCMC-sampling and to avoid density estimates from MCMC-samples \citep{mehrabi_information_2019}.

Similarly, the expectation value of $\chi^2(y|\theta)/2+\phi(\theta)$ can be computed directly with the derivative
\begin{equation}
\frac{\partial}{\partial\beta}\ln Z[\beta]\Bigg|_{\beta = 1} = 
-\left\bra\frac{1}{2}\chi^2(y|\theta) + \phi(\theta)\right\ket = 
-\frac{1}{2}F_{\mu\nu} F^{\mu\nu} = -\frac{n}{2},
\end{equation}
where the result applies to a simple quadratic model $\chi^2 = F_{\alpha\beta}\theta^\alpha\theta^\beta$ and for a flat prior $\phi$, using the fact that the covariance $\bra\theta^\mu\theta^\nu\ket = F^{\mu\nu}$ and the Fisher-matrix $F_{\mu\nu}$ are inverse, such that $F_{\mu\nu}F^{\mu\nu} = \delta_\mu^\mu = n$. Additionally, the variance around that particular value follows from
\begin{equation}
\frac{\partial^2}{\partial\beta^2}\ln Z[\beta]\Bigg|_{\beta = 1} = 
\left\bra\left(\frac{1}{2}\chi^2(y|\theta) + \phi(\theta)\right)^2\right\ket - \left\bra\frac{1}{2}\chi^2(y|\theta) + \phi(\theta)\right\ket^2 = 
\frac{1}{4}\left(F_{\mu\nu}F^{\mu\nu}F_{\rho\sigma}F^{\rho\sigma} + F_{\mu\nu}F^{\mu\rho}F_{\rho\sigma}F^{\sigma\nu}-F_{\mu\nu}F^{\mu\sigma}F_{\rho\sigma}F^{\nu\rho}\right) -\frac{n^2}{4}=
\frac{n}{2}.
\end{equation}
where the forth moment in the expectation value of $\bra(\chi^2/2)^2\ket$ was decomposed with the Wick-theorem, giving rise to three terms. These two quantities generalise to higher derivative orders and could potentially be diagnostic tools for the convergence of Markov-chains.

\subsection{Hamilton Monte-Carlo partitions}
Extensions to the partition function $Z[\beta,J_\alpha]$ could consist of including a kinetic term
\begin{equation}
Z[\beta,J_\alpha] = 
\int\frac{\dd^n p}{(2\pi)^n}\:\int\dd^n\theta\:
\exp\left(-\beta\left[\frac{p^2}{2m} + \frac{1}{2}\chi^2(y|\theta) + \phi(\theta)\right]\right)\exp(\beta J_\alpha\theta^\alpha) =
\int\frac{\dd^n p}{(2\pi)^n}\:\int\dd^n\theta\:
\exp\left(-\beta\mathcal{H}(p,\theta)\right)\exp(\beta J_\alpha\theta^\alpha)
\end{equation}
which would correspond to Hamilton-Monte Carlo Markov-chains, with a Hamilton function
\begin{equation}
\mathcal{H}(p,\theta) = \frac{p^2}{2m} + \frac{1}{2}\chi^2(y|\theta) + \phi(\theta)
\end{equation}
not only reflecting potential, but also kinetic energy \citep[for an application in cosmology, see][]{jasche_fast_2010, kitaura_bayesian_2008}. The integrals in the partition sum separate always, such that there is no influence of the $\dd^np$-integration on the cumulants of $p(\theta|y)$, explicitly for instance for a linear model with a flat prior as discussed in eqn.~(\ref{eqn_linear_model})
\begin{equation}
Z[\beta,J_\alpha] = 
\int\frac{\dd^n p}{(2\pi)^n}\:\int\dd^n\theta\:
\exp\left(-\beta\left[\frac{p^2}{2m} + \frac{1}{2}\chi^2(y|\theta) + \phi(\theta)\right]\right)\exp(\beta J_\alpha\theta^\alpha) = 
\sqrt{\frac{m^n}{\beta^{2n}}\frac{1}{\mathrm{det}(F)}}\exp\left(\frac{\beta}{2}F^{\alpha\beta}J_\alpha J_\beta\right),
\end{equation}
however, the inclusion of momentum degrees of freedom can make the sampling more efficient, in particular in the case of strong degeneracies.

\subsection{Ensemble Monte-Carlo partitions}
Having $N$ instead of a single Monte-Carlo Markov-chain bridges towards ensemble Markov-methods \citep{foreman-mackey_emcee:_2013}: If the chains are non-interacting, this would amount to a factorising $N$-particle partition $Z[\beta,J_\mu,N]$
\begin{equation}
Z[\beta,J_\alpha,N] = Z[\beta,J_\alpha]^N
\end{equation}
suggesting the question whether one could introduce a fugacity $z$ that controls the number $N$ of Markov-chains, effectively defining a macrocanonical partition
\begin{equation}
Z[\beta,J_\alpha,z] = 
\sum_N Z[\beta,J_\alpha,N] z^N = 
\sum_N \left(z\: Z[\beta,J_\alpha]\right)^N.
\end{equation}
Adding a Gibbs-factor $1/N!$ seems superfluous as the chains are perfectly distinguishable.

\section{application to supernova data}
\label{sect_supernova}
As a topical example for a non-Gaussian likelihood from cosmology we consider constraints on $\Omega_m$ and $w$ from the distance redshift relation of supernovae \citep{Riess1998, goobar_supernova_2011}, where we focus on spatially flat FLRW-cosmologies with a constant dark energy equation of state, and derive constraints on $\Omega_m$ and $w$ for the Union2.1-data set \citep{suzuki_hubble_2012, amanullah_spectra_2010, kowalski_improved_2008}. For these cases, the distance modulus $y(z)$ as a function of redshift $z$ is given by
\begin{equation} \label{eqn:DistModulus}
y(z,\Omega_m,w) = 10 + 5 \log\left((1+z) \, \chi_H \int_{0}^{z}\dd{z^\prime}\:
\frac{1}{\sqrt{\Omega_m (1+z^\prime)^3 + (1-\Omega_m) (1+z^\prime)^{3(1+w)}}}\right),
\end{equation}
with the relevant integral, for this type of cosmology, being expressible in terms of a hypergeometric function ${}_{2}F_{1}$ \citep{rafael_paper}
\begin{equation}
\int\dd{u}\:\frac{1}{\sqrt{A \, u^3 + B \, u^c}} = 
-\frac{2 u \sqrt{\frac{A \, u^{3 - c}}{B} + 1} \, {}_{2}F_{1}\left(\frac{1}{2}, \frac{c - 2}{2c - 6}; \frac{3 c - 8}{2 c - 6}; -\frac{A \, u^{3 - c}}{B}\right)}{(c - 2) \sqrt{A \, u^3 + B \, u^c}} + \text{const.}
\end{equation}
Expressing the likelihood for the two parameters $\Omega_m$ and $w$ for Gaussian errors $\sigma_i$ in the distance moduli $y_i$ yields a simplified expression, where we neglect correlations between the data points,
\begin{equation}
\likeli(y|\Omega_m,w) \propto \exp\left(-\frac{1}{2}\chi^2(y|\Omega_m,w)\right)
\quad\text{with}\quad
\chi^2(y|\Omega_m,w) = \sum_i\left(\frac{y_i - y(z_i,\Omega_m,w)}{\sigma_i}\right)^2,
\label{eqn:MCMC_likelihood}
\end{equation}
which we use for the construction of partition functions $Z[\beta,J_\alpha]$ and for implementation in a Monte-Carlo Markov-chain. For simplicity, we employ a flat prior $\pi(\Omega_m,w)$ on the two cosmological parameters and drive the MCMC-sampling with a straightforward Metropolis-Hastings algorithm. The model for the distance modulus as a function of the model parameters $\Omega_m$ and $w$ is clearly nonlinear, giving rise to a non-Gaussian likelihood, on which we demonstrate the a Gram-Charlier-type expansion, with cumulants derived from differentiation of the partition function $\ln Z[\beta,J_\alpha]$. These differentiations are done by finite differencing of a numerically integrated partition function, where we have verified the numerical precision in toy examples.

In the case that the partition function cannot be computed exactly or be approximated analytically it can be constructed from a Monte-Carlo Markov-chain sampling from a posterior distribution by computing an expectation value of $\exp \left(J_\alpha \theta^\alpha\right)$ as
\begin{align}
Z[\beta=1, J_\alpha] = 
\left<\exp \left(J_\alpha \theta^\alpha\right)\right>\approx
\frac{1}{N}\sum_{i=1}^{N} \exp \left(J_\alpha(\theta^\alpha)_i\right).
\end{align} 
Here, $i$ denotes the $i$-th sample generated by the Metropolis-Hastings algorithm. This allows to find the cumulants following the first part of eqn.~(\ref{eqn:cumulants_nG}) and to construct the partition function for a given $J_\alpha$. The cumulants can then be computed through finite differencing with respect to all $J_\alpha$.
	
The numerical precision can be verified by computing moments of order $a+b$ of the posterior using the samples drawn by the Monte-Carlo Markov-chain, which are distributed according to the posterior distribution. They are computed as 
\begin{align}
\left<\Omega\indices{_m^a} w^b\right> = 
\int\dd\Omega_m\dd w\: p(\Omega_m, w|y)\:\Omega\indices{_m^a} w^b \approx 
\frac{1}{N}\sum_{i=1}^{N} (\Omega\indices{_m^a} w^b)_i,
\end{align}
directly from the samples, with a subsequent conversion of moments to cumulants with Faa di Bruno's formula. 

\begin{figure}
		\centering
		\includegraphics[width =0.45 \textwidth]{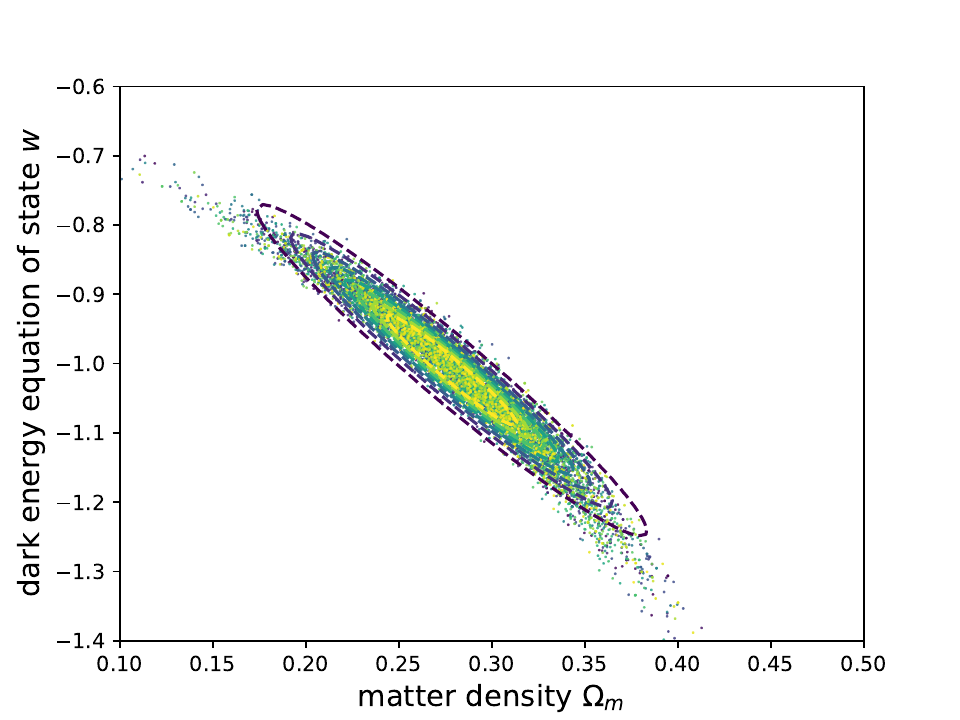}
		\includegraphics[width =0.45 \textwidth]{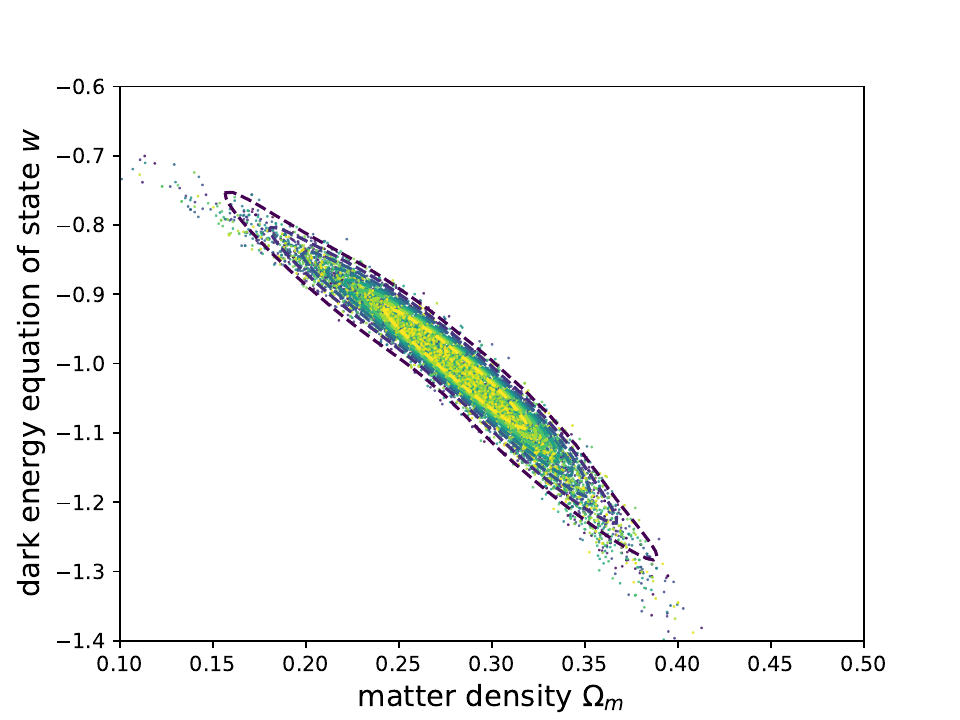}
\caption{The scatter plot depicts $10^4$ points from a Monte-Carlo Markov-chain with a likelihood as described in eqn.~(\ref{eqn:MCMC_likelihood}). The superimposed contours are the 0th (left) and 1st (right) order approximations to this posterior distribution using the a Fisher-matrix (left) and Gram-Charlier series with nonzero skewness (right).}
\label{fig:supernova_approx}
\end{figure}

At Gaussian and lowest non-Gaussian order, posterior distribution $p(\Omega_m,w|y)$ is depicted in Fig.~ \ref{fig:supernova_approx} along with the samples generated by the Metropolis-Hastings algorithm. The Gaussian isoprobability contours correspond exactly to the Fisher-matrix, and the lowest non-Gaussian approximation to a Gram-Charlier expansion including skewness. Driving the Gram-Charlier-expansion to higher order shows the known deficiency to reproduce distributions with strong non-Gaussianties, causing the Gram-Charlier-expansion lose positive definiteness; those are cases where DALI plays its unique strength \citep{sellentin_fast_2015, sellentin_breaking_2014}. We would like to emphasise though that the computation of the cumulants from $\ln Z$ is numerically sound.
	
In fact, the computation of cumulants of the posterior distribution from the two ways considered in this work give very similar results. Table~\ref{table_comparison} summarises all cumulants up to forth order from the posterior distribution $p(\Omega_m,w|y)$ of the supernova example, where these cumulants follow either from estimation of the moments from MCMC-samples and successive conversion into cumulants using Faa di Bruno's formula, or by finite differencing of the numerically evaluated partition function $\ln Z$. The numerical experiment suggests that both methods yield very similar results. Note however that both of these results are entirely dependent on the same part of a Markov-chain. Comparing the results in table \ref{table_comparison} to cumulants obtained from three different parts of the same Markov-chain, each containing $10^6$ elements lead to differences in the cumulant values of about $0.07\%$, $0.4\%$, $20\%$, $20\%$, for the first, second, third and fourth cumulant respectively. We would like to emphasise that given the large number of samples needed, we replaced the Metropolis-Hasings algorithm by the affine-invariant sampler emcee \citep{foreman-mackey_emcee:_2013} for better efficiency.

\begin{table}
	\centering
	\begin{tabular}{lrr}
		\hline\hline
		cumulants & MCMC-samling & partition function\\
		\hline
		$\kappa^{\Omega_m}$&$0.27881995$&$0.27882000$\\
		$\kappa^{w}$&$-1.01051052$&$-1.01051010$\\
		\hline
		$\kappa^{\Omega_m,\Omega_m}$&$0.0021031685$&$0.0021031674$\\
		$\kappa^{\Omega_m,w}$&$-0.004649421$&$-0.004649417$\\
		$\kappa^{w,w}$&$0.010934010$&$0.010934001$\\
		\hline
		$\kappa^{\Omega_m,\Omega_m,\Omega_m}$&$-3.565244\cdot 10^{-5}$&$-3.565206\cdot 10^{-5}$\\
		$\kappa^{\Omega_m,\Omega_m,w}$&$3.7907932\cdot 10^{-5}$&$3.7907907\cdot 10^{-5}$\\
		$\kappa^{\Omega_m,w,w}$&$1.082203\cdot 10^{-5}$&$1.082211\cdot 10^{-5}$\\
		$\kappa^{w,w,w}$&$-0.000250971$&$-0.000250968$\\
		\hline
		$\kappa^{\Omega_m,\Omega_m,\Omega_m,\Omega_m}$&$1.2953\cdot 10^{-6}$&$1.2961\cdot 10^{-6}$\\
		$\kappa^{\Omega_m,\Omega_m,\Omega_m,w}$&$-1.23547\cdot 10^{-6}$&$-1.23531\cdot 10^{-6}$\\
		$\kappa^{\Omega_m,\Omega_m,w,w}$&$7.911\cdot 10^{-7}$&$7.906\cdot 10^{-7}$\\
		$\kappa^{\Omega_m,w,w,w}$&$-1.27243\cdot 10^{-6}$&$-1.27230\cdot 10^{-6}$\\
		$\kappa^{w,w,w,w}$&$1.1046275\cdot 10^{-5}$&$1.1046293\cdot 10^{-5}$\\
		\hline
	\end{tabular}
	\caption{Comparison of the cumulants $\kappa$ of order 1, 2, 3 and 4 of the supernova posterior distribution $p(\Omega_m,w|y)$, evaluated by MCMC-sampling and by finite differencing of the logarithmic partition function $\ln Z$. The cumulants are computed from a Markov chain with $10^6$ elements.}
	\label{table_comparison}
\end{table}

\section{summary and discussion}
\label{sect_summary}
The subject of this paper was an analytical approach to inference in cosmology with Gaussian and weakly non-Gaussian distributions, with supernova constraints on $\Omega_m$ and $w$ as a topical example from cosmology with a weakly non-Gaussian likelihood. Our motivation was the statistical model behind Monte-Carlo Markov-chain methods routinely used in likelihood evaluation with the hope to develop analytical methods that extend the Fisher-matrix formalism beyond second order and that allow an approximative description of non-Gaussian posterior distributions.

\begin{enumerate}
\item{It is a known result how the Bayes-evidence $p(y)$ can be extended to a partition function $Z[\beta,J_\alpha]$, by introducing an inverse temperature $\beta$ and by carrying out a Laplace-transform of the actual model parameter $\theta^\alpha$ to source variables $J_\alpha$. Cumulants of the posterior distribution $p(\theta|y)$ of order $n$ would follow from $n$-fold differentiation of the logarithmic partition function. An MCMC-algorithm would sample microstates from this partition function, so that the determination of cumulants of order $n$ would be possible from estimating all moments up to order $n$ and assembling them according to Faa di Bruno's formula: Instead, we aim to compute the integral defining the partition function directly, numerical if necessary, and determine the cumulants by finite differencing.}

\item{We recover analytically the well-known Fisher-matrix formalism for linear models from our approach. The partition-function is of Gaussian shape and has an analytical solution, again of Gaussian form. The sequence of cumulants truncates at second order such that one obtains a Gaussian posterior distribution $p(\theta|y)$. While the first cumulant becomes equal to the true model parameters as expected due to the Gau{\ss}-Markov theorem, the second cumulant (or the covariance) is given by the inverse Fisher-matrix. We were able to show within the partition function-formalism by functional extremisation the Cram{\'e}r-Rao inequality, i.e. that the inverse Fisher-matrix is in fact a lower bound to the variance and that this bound is reached for a $\chi^2$ that is quadratic in the paramters, or equivalently, a linear model.}

\item{The partition function gives a direct access to higher-order cumulants and Gram-Charlier expansions of the posterior distribution. Cumulants of arbitrary order $n$ follow by $n$-fold differentiation, and we have validated the numerics in comparison to MCMC-sampling in toy-examples. More importantly, though, is the factorisation of the partition function of a nonlinear model into a Gaussian and non-Gaussian part, where in the case of weak non-Gaussianities the corresponding term can be Taylor-expanded, leading straight to the Gram-Charlier expansion of weakly non-Gaussian distributions.}

\item{It is straightforward to show that derivatives of order $n$ of the logarithmic partition function with respect to inverse temperature $\beta$ correspond to cumulants of the $\chi^2(y|\theta)$-function. Motivated by the result from statistical mechanics, where one defines the Helmholtz free energy $F(\beta)$ as $-\ln Z/\beta$ and where the entropy $S$ is given by $\beta^2\partial F/\partial\beta^2$ we showed that the Shannon-entropy of the posterior distribution $p(\theta|y)$ follows analogously. Extension of the partition function to include kinetic terms would lead to Hamilton Monte-Carlo partitions. We showed that quadratic kinetic terms would not change the value of cumulants of the posterior distribution.}

\item{Application of our formalism to supernova data showed that the posterior distribution $p(\Omega_m,w|y)$ for the two cosmological parameters $\Omega_m$ and $w$ can be evaluated easily to fourth order in the cumulants by finite differentiation of $\ln Z$, and that the cumulants correspond well numerically to estimates from samples obtained in MCMC-sampling.}

\end{enumerate}

We intend to pursue the idea of ensemble MCMC-methods controlled by a fugacity $z$ further, and hope to find methods to evaluate Bayes-evidences in the partition function formalism, as well as bridge to measures of statistical consistency \citep{raveri_quantifying_2020, raveri_concordance_2019}. Changes in fugacity $z$ as a state variable would yield a pathway of controlling the sampling analogous to changing $\beta$ in simulated annealing.

\section*{acknowledgements}
This work was supported by the Deutsche Forschungsgemeinschaft (DFG, German Research Foundation) under
Germany's Excellence Strategy EXC 2181/1 - 390900948 (the Heidelberg STRUCTURES Excellence Cluster). We would like to thank $\ln(a)$ Sellentin for thoughtful remarks concerning DALI.

\section*{data availability statement}
Our python-toolkit for the automatic computation of multivariate cumulants and the corresponding Gram-Charlier-expansion is available on request.

\bibliographystyle{mnras}
\bibliography{references}

\begin{thebibliography}{}
\makeatletter
\relax
\def\mn@urlcharsother{\let\do\@makeother \do\$\do\&\do\#\do\^\do\_\do\%\do\~}
\def\mn@doi{\begingroup\mn@urlcharsother \@ifnextchar [ {\mn@doi@}
  {\mn@doi@[]}}
\def\mn@doi@[#1]#2{\def\@tempa{#1}\ifx\@tempa\@empty \href
  {http://dx.doi.org/#2} {doi:#2}\else \href {http://dx.doi.org/#2} {#1}\fi
  \endgroup}
\def\mn@eprint#1#2{\mn@eprint@#1:#2::\@nil}
\def\mn@eprint@arXiv#1{\href {http://arxiv.org/abs/#1} {{\tt arXiv:#1}}}
\def\mn@eprint@dblp#1{\href {http://dblp.uni-trier.de/rec/bibtex/#1.xml}
  {dblp:#1}}
\def\mn@eprint@#1:#2:#3:#4\@nil{\def\@tempa {#1}\def\@tempb {#2}\def\@tempc
  {#3}\ifx \@tempc \@empty \let \@tempc \@tempb \let \@tempb \@tempa \fi \ifx
  \@tempb \@empty \def\@tempb {arXiv}\fi \@ifundefined
  {mn@eprint@\@tempb}{\@tempb:\@tempc}{\expandafter \expandafter \csname
  mn@eprint@\@tempb\endcsname \expandafter{\@tempc}}}

\bibitem[\protect\citeauthoryear{Amanullah et~al.,}{Amanullah
  et~al.}{2010}]{amanullah_spectra_2010}
Amanullah R.,  et~al., 2010, \mn@doi [ApJ] {10.1088/0004-637X/716/1/712}, 716,
  712

\bibitem[\protect\citeauthoryear{Amara \& Kitching}{Amara \&
  Kitching}{2011}]{amara_figures_2011}
Amara A.,  Kitching T.~D.,  2011, \mn@doi [MNRAS]
  {10.1111/j.1365-2966.2010.17947.x}, 413, 1505

\bibitem[\protect\citeauthoryear{Amara \& Refregier}{Amara \&
  Refregier}{2007}]{amara_systematic_2007}
Amara A.,  Refregier A.,  2007, arXiv e-prints 0710.5171, 710

\bibitem[\protect\citeauthoryear{Amari}{Amari}{2016}]{amari_information_2016}
Amari S.-i.,  2016, Information Geometry and Its Applications.
 Applied Mathematical Sciences Vol. 194, Springer Japan,
  \mn@doi{10.1007/978-4-431-55978-8}, \url
  {http://link.springer.com/10.1007/978-4-431-55978-8}

\bibitem[\protect\citeauthoryear{Amendola \& Tsujikawa}{Amendola \&
  Tsujikawa}{2010}]{AmendolaDarkEnergy}
Amendola L.,  Tsujikawa S.,  2010, Dark Energy.
Camebridge University Press, Cambridge, MA

\bibitem[\protect\citeauthoryear{{Arutjunjan}, {Sch{\"a}fer}  \&
  {Kreutz}}{{Arutjunjan} et~al.}{2022}]{rafael_paper}
{Arutjunjan} R.,  {Sch{\"a}fer} B.~M.,   {Kreutz} C.,  2022, to be submitted to
  JRSSB

\bibitem[\protect\citeauthoryear{Baez \& Fritz}{Baez \&
  Fritz}{2014}]{baez_bayesian_2014}
Baez J.~C.,  Fritz T.,  2014, arXiv e-prints 1402.3067

\bibitem[\protect\citeauthoryear{Bassett, Fantaye, Hlozek  \& Kotze}{Bassett
  et~al.}{2009}]{bassett_fisher4cast_2009}
Bassett B.~A.,  Fantaye Y.,  Hlozek R.,   Kotze J.,  2009, arXiv e-prints
  0906.0974

\bibitem[\protect\citeauthoryear{Bassett, Fantaye, Hlozek  \& Kotze}{Bassett
  et~al.}{2011}]{bassett_fisher_2011}
Bassett B.~A.,  Fantaye Y.,  Hlozek R.,   Kotze J.,  2011, International
  Journal of Modern Physics D, 20, 2559

\bibitem[\protect\citeauthoryear{Bellini, Cuesta, Jimenez  \& Verde}{Bellini
  et~al.}{2016}]{bellini_constraints_2016}
Bellini E.,  Cuesta A.~J.,  Jimenez R.,   Verde L.,  2016, \mn@doi [JCAP]
  {10.1088/1475-7516/2016/02/053}, 1602, 053

\bibitem[\protect\citeauthoryear{{Berkowitz} \& {Garner}}{{Berkowitz} \&
  {Garner}}{1970}]{multivariate_gc}
{Berkowitz} S.,  {Garner} F.~J.,  1970, Mathematics of Computation, \href
  {https://ui.adsabs.harvard.edu/abs/1995AmJPh..63..620A} {24, 537}

\bibitem[\protect\citeauthoryear{Carron, Amara  \& Lilly}{Carron
  et~al.}{2011}]{carron_probe_2011}
Carron J.,  Amara A.,   Lilly S.,  2011, arXiv e-prints 1107.0726

\bibitem[\protect\citeauthoryear{Chung, Everett  \& Riotto}{Chung
  et~al.}{2003}]{Chung2003}
Chung D.,  Everett L.,   Riotto A.,  2003, \mn@doi [Physics Letters B]
  {https://doi.org/10.1016/S0370-2693(03)00099-6}, 556, 61

\bibitem[\protect\citeauthoryear{Coe}{Coe}{2009}]{coe_fisher_2009}
Coe D.,  2009, arXiv e-prints 0906.4123

\bibitem[\protect\citeauthoryear{Cram{\'e}r}{Cram{\'e}r}{1999}]{cramer1999mathematical}
Cram{\'e}r H.,  1999, Mathematical methods of statistics.
Princeton university press

\bibitem[\protect\citeauthoryear{{Crooks}, {Dunn}, {Frampton}  \&
  {Ng}}{{Crooks} et~al.}{2000}]{QuintessenceCMB2}
{Crooks} J.~L.,  {Dunn} J.~O.,  {Frampton} P.~H.,   {Ng} Y.~J.,  2000, arXiv
  e-prints, \href {https://ui.adsabs.harvard.edu/abs/2000astro.ph..5406C} {pp
  astro--ph/0005406}

\bibitem[\protect\citeauthoryear{Elsner \& Wandelt}{Elsner \&
  Wandelt}{2012}]{elsner_fast_2012}
Elsner F.,  Wandelt B.~D.,  2012, \mn@doi [Astronomy \& Astrophysicas]
  {10.1051/0004-6361/201218985}, 540, L6

\bibitem[\protect\citeauthoryear{Foreman-Mackey, Hogg, Lang  \&
  Goodman}{Foreman-Mackey et~al.}{2013}]{foreman-mackey_emcee:_2013}
Foreman-Mackey D.,  Hogg D.~W.,  Lang D.,   Goodman J.,  2013, \mn@doi
  [Publications of the Astronomical Society of the Pacific] {10.1086/670067},
  125, 306

\bibitem[\protect\citeauthoryear{Giesel, Reischke, Sch{\"a}fer  \& Chia}{Giesel
  et~al.}{2021}]{giesel_information_2021}
Giesel E.,  Reischke R.,  Sch{\"a}fer B.~M.,   Chia D.,  2021, \mn@doi [JCAP]
  {10.1088/1475-7516/2021/01/005}, 2021, 005

\bibitem[\protect\citeauthoryear{Goobar \& Leibundgut}{Goobar \&
  Leibundgut}{2011}]{goobar_supernova_2011}
Goobar A.,  Leibundgut B.,  2011, \mn@doi [Annual Review of Nuclear and
  Particle Science] {10.1146/annurev-nucl-102010-130434}, 61, 251

\bibitem[\protect\citeauthoryear{Grandis, Seehars, Refregier, Amara  \&
  Nicola}{Grandis et~al.}{2016}]{grandis_information_2016}
Grandis S.,  Seehars S.,  Refregier A.,  Amara A.,   Nicola A.,  2016, \mn@doi
  [JCAP] {10.1088/1475-7516/2016/05/034}, 2016, 034

\bibitem[\protect\citeauthoryear{Grassi, Heisenberg, Byrnes  \&
  Schaefer}{Grassi et~al.}{2014}]{grassi_test_2014}
Grassi A.,  Heisenberg L.,  Byrnes C.~T.,   Schaefer B.~M.,  2014, \mn@doi
  [MNRAS] {10.1093/mnras/stu900}, 442, 1068

\bibitem[\protect\citeauthoryear{Handley \& Lemos}{Handley \&
  Lemos}{2019}]{handley_quantifying_2019}
Handley W.,  Lemos P.,  2019, arXiv e-prints 1903.06682

\bibitem[\protect\citeauthoryear{Jasche \& Kitaura}{Jasche \&
  Kitaura}{2010}]{jasche_fast_2010}
Jasche J.,  Kitaura F.~S.,  2010, \mn@doi [MNRAS]
  {10.1111/j.1365-2966.2010.16897.x}, 407, 29

\bibitem[\protect\citeauthoryear{Jaynes}{Jaynes}{1957}]{jaynes_information_1957}
Jaynes E.~T.,  1957, \mn@doi [Physical Review] {10.1103/PhysRev.106.620}, 106,
  620

\bibitem[\protect\citeauthoryear{Jenkins \& Peacock}{Jenkins \&
  Peacock}{2011}]{jenkins_power_2011}
Jenkins C.~R.,  Peacock J.~A.,  2011, \mn@doi [MNRAS]
  {10.1111/j.1365-2966.2011.18361.x}, 413, 2895

\bibitem[\protect\citeauthoryear{Johnson}{Johnson}{2002}]{johnson_curious_2002}
Johnson W.~P.,  2002, \mn@doi [The American Mathematical Monthly]
  {10.2307/2695352}, 109, 217

\bibitem[\protect\citeauthoryear{Joudaki et~al.,}{Joudaki
  et~al.}{2017}]{joudaki_cfhtlens_2017}
Joudaki S.,  et~al., 2017, \mn@doi [MNRAS] {10.1093/mnras/stw2665}, 465, 2033

\bibitem[\protect\citeauthoryear{{Juszkiewicz}, {Weinberg}, {Amsterdamski},
  {Chodorowski}  \& {Bouchet}}{{Juszkiewicz}
  et~al.}{1995}]{1995ApJ...442...39J}
{Juszkiewicz} R.,  {Weinberg} D.~H.,  {Amsterdamski} P.,  {Chodorowski} M.,
  {Bouchet} F.,  1995, \mn@doi [ApJ] {10.1086/175420}, \href
  {https://ui.adsabs.harvard.edu/abs/1995ApJ...442...39J} {442, 39}

\bibitem[\protect\citeauthoryear{Kerscher \& Weller}{Kerscher \&
  Weller}{2019}]{kerscher_model_2019}
Kerscher M.,  Weller J.,  2019, \mn@doi [SciPost]
  {10.21468/SciPostPhysLectNotes.9}, p.~9

\bibitem[\protect\citeauthoryear{Kitaura \& Ensslin}{Kitaura \&
  Ensslin}{2008}]{kitaura_bayesian_2008}
Kitaura F.~S.,  Ensslin T.~A.,  2008, \mn@doi [MNRAS]
  {10.1111/j.1365-2966.2008.13341.x}, 389, 497

\bibitem[\protect\citeauthoryear{Kitching \& Taylor}{Kitching \&
  Taylor}{2011}]{kitching_path_2011}
Kitching T.~D.,  Taylor A.~N.,  2011, \mn@doi [MNRAS]
  {10.1111/j.1365-2966.2010.17548.x}, 410, 1677

\bibitem[\protect\citeauthoryear{Kitching, Amara, Abdalla, Joachimi  \&
  Refregier}{Kitching et~al.}{2009}]{kitching_cosmological_2009}
Kitching T.~D.,  Amara A.,  Abdalla F.~B.,  Joachimi B.,   Refregier A.,  2009,
  \mn@doi [MNRAS] {10.1111/j.1365-2966.2009.15408.x}, 399, 2107

\bibitem[\protect\citeauthoryear{Knuth, Habeck, Malakar, Mubeen  \&
  Placek}{Knuth et~al.}{2015}]{knuth_bayesian_2015}
Knuth K.~H.,  Habeck M.,  Malakar N.~K.,  Mubeen A.~M.,   Placek B.,  2015,
  \mn@doi [Digital Signal Processing] {10.1016/j.dsp.2015.06.012}, 47, 50

\bibitem[\protect\citeauthoryear{Kowalski et~al.,}{Kowalski
  et~al.}{2008}]{kowalski_improved_2008}
Kowalski M.,  et~al., 2008, \mn@doi [ApJ] {10.1086/589937}, 686, 749

\bibitem[\protect\citeauthoryear{Lewis \& Bridle}{Lewis \&
  Bridle}{2002}]{lewis_cosmological_2002}
Lewis A.,  Bridle S.,  2002, \mn@doi [PRD] {10.1103/PhysRevD.66.103511}, 66

\bibitem[\protect\citeauthoryear{Liddle, Mukherjee, Parkinson  \& Wang}{Liddle
  et~al.}{2006}]{liddle_present_2006}
Liddle A.~R.,  Mukherjee P.,  Parkinson D.,   Wang Y.,  2006, \mn@doi [PRD]
  {10.1103/PhysRevD.74.123506}, 74

\bibitem[\protect\citeauthoryear{Loverde, Hui  \& Gazta{\~n}aga}{Loverde
  et~al.}{2007}]{loverde_magnification-temperature_2007}
Loverde M.,  Hui L.,   Gazta{\~n}aga E.,  2007, \mn@doi [PRD]
  {10.1103/PhysRevD.75.043519}, 75, 043519

\bibitem[\protect\citeauthoryear{Mehrabi \& Ahmadi}{Mehrabi \&
  Ahmadi}{2019}]{mehrabi_information_2019}
Mehrabi A.,  Ahmadi A.,  2019, arXiv e-prints 1904.11920

\bibitem[\protect\citeauthoryear{Metropolis}{Metropolis}{1985}]{metropolis_monte-carlo:_1985}
Metropolis N.,  1985, in Alcouffe R.,  Dautray R.,  Forster A.,  Ledonois G.,
  Mercier B.,  eds,  Lecture Notes in Physics, Berlin Springer Verlag Vol. 240,
  Lecture Notes in Physics, Berlin Springer Verlag. p.~62,
  \mn@doi{10.1007/BFb0049035}

\bibitem[\protect\citeauthoryear{Metropolis, Rosenbluth, Rosenbluth, Teller  \&
  Teller}{Metropolis et~al.}{1953}]{metropolis_equation_1953}
Metropolis N.,  Rosenbluth A.~W.,  Rosenbluth M.~N.,  Teller A.~H.,   Teller
  E.,  1953, \mn@doi [The Journal of Chemical Physics] {10.1063/1.1699114}, 21,
  1087

\bibitem[\protect\citeauthoryear{Mortonson, Weinberg  \& White}{Mortonson
  et~al.}{2013}]{mortonson2013dark}
Mortonson M.~J.,  Weinberg D.~H.,   White M.,  2013, Dark Energy: A Short
  Review (\mn@eprint {arXiv} {1401.0046})

\bibitem[\protect\citeauthoryear{Nicola, Amara  \& Refregier}{Nicola
  et~al.}{2018}]{nicola_consistency_2018}
Nicola A.,  Amara A.,   Refregier A.,  2018, arXiv e-prints 1809.07333

\bibitem[\protect\citeauthoryear{Perlmutter \& {et al.}}{Perlmutter \& {et
  al.}}{2003}]{perlmutter_supernovae_2003}
Perlmutter S.,  {et al.} 2003, Physics today, 56, 53

\bibitem[\protect\citeauthoryear{Pinho, Reischke, Teich  \& Sch\"afer}{Pinho
  et~al.}{2021}]{Pinho:2020uzv}
Pinho A.~M.,  Reischke R.,  Teich M.,   Sch\"afer B.~M.,  2021, \mn@doi [MNRAS]
  {10.1093/mnras/stab561}, 503, 1187

\bibitem[\protect\citeauthoryear{Raveri \& Hu}{Raveri \&
  Hu}{2019}]{raveri_concordance_2019}
Raveri M.,  Hu W.,  2019, \mn@doi [PRD] {10.1103/PhysRevD.99.043506}, 99,
  043506

\bibitem[\protect\citeauthoryear{Raveri, Martinelli, Zhao  \& Wang}{Raveri
  et~al.}{2016}]{raveri_cosmicfish_2016}
Raveri M.,  Martinelli M.,  Zhao G.,   Wang Y.,  2016, \mn@doi [arXiv e-prints
  11606.06268] {10.48550/arXiv.1606.06268}

\bibitem[\protect\citeauthoryear{Raveri, Zacharegkas  \& Hu}{Raveri
  et~al.}{2020}]{raveri_quantifying_2020}
Raveri M.,  Zacharegkas G.,   Hu W.,  2020, \mn@doi [PRD]
  {10.1103/PhysRevD.101.103527}, 101, 103527

\bibitem[\protect\citeauthoryear{Refregier, Amara, Kitching  \&
  Rassat}{Refregier et~al.}{2011}]{refregier_icosmo:_2011}
Refregier A.,  Amara A.,  Kitching T.~D.,   Rassat A.,  2011, \mn@doi [AAP]
  {10.1051/0004-6361/200811112}, 528, A33+

\bibitem[\protect\citeauthoryear{Reischke, Kiessling  \& Sch{\"a}fer}{Reischke
  et~al.}{2017}]{reischke_variations_2016}
Reischke R.,  Kiessling A.,   Sch{\"a}fer B.~M.,  2017, MNRAS, 465, 4016

\bibitem[\protect\citeauthoryear{Riess et~al.,}{Riess et~al.}{1998}]{Riess1998}
Riess A.~G.,  et~al., 1998, The Astronomical Journal, 116, 1009

\bibitem[\protect\citeauthoryear{Sch{\"a}fer \& Heisenberg}{Sch{\"a}fer \&
  Heisenberg}{2012}]{schafer_weak_2012}
Sch{\"a}fer B.~M.,  Heisenberg L.,  2012, \mn@doi [MNRAS]
  {10.1111/j.1365-2966.2012.21137.x}, 423, 3445

\bibitem[\protect\citeauthoryear{Sch{\"a}fer \& Reischke}{Sch{\"a}fer \&
  Reischke}{2016}]{schafer_describing_2016}
Sch{\"a}fer B.~M.,  Reischke R.,  2016, \mn@doi [MNRAS]
  {10.1093/mnras/stw1221}, 460, 3398

\bibitem[\protect\citeauthoryear{Sellentin}{Sellentin}{2015}]{sellentin_fast_2015}
Sellentin E.,  2015, \mn@doi [MNRAS] {10.1093/mnras/stv1671}, 453, 893

\bibitem[\protect\citeauthoryear{Sellentin, Quartin  \& Amendola}{Sellentin
  et~al.}{2014}]{sellentin_breaking_2014}
Sellentin E.,  Quartin M.,   Amendola L.,  2014, \mn@doi [MNRAS]
  {10.1093/mnras/stu689}, 441, 1831

\bibitem[\protect\citeauthoryear{Suzuki et~al.,}{Suzuki
  et~al.}{2012}]{suzuki_hubble_2012}
Suzuki N.,  et~al., 2012, ApJ, 746, 85

\bibitem[\protect\citeauthoryear{Taburet, Aghanim, Douspis  \& Langer}{Taburet
  et~al.}{2009}]{taburet_biases_2009}
Taburet N.,  Aghanim N.,  Douspis M.,   Langer M.,  2009, \mn@doi [MNRAS]
  {10.1111/j.1365-2966.2008.14105.x}, 392, 1153

\bibitem[\protect\citeauthoryear{Tegmark, Taylor  \& Heavens}{Tegmark
  et~al.}{1997}]{tegmark_karhunen-loeve_1997}
Tegmark M.,  Taylor A.,   Heavens A.,  1997, \mn@doi [The Astrophysical
  Journal] {10.1086/303939}, 480, 22

\bibitem[\protect\citeauthoryear{Trotta}{Trotta}{2007}]{trotta_applications_2007}
Trotta R.,  2007, \mn@doi [MNRAS] {10.1111/j.1365-2966.2007.11738.x}, 378, 72

\bibitem[\protect\citeauthoryear{Trotta}{Trotta}{2008}]{trotta_bayes_2008}
Trotta R.,  2008, \mn@doi [Contemporary Physics] {10.1080/00107510802066753},
  49, 71

\bibitem[\protect\citeauthoryear{Trotta}{Trotta}{2017}]{trotta_bayesian_2017}
Trotta R.,  2017, arXiv e-prints 1701.01467

\bibitem[\protect\citeauthoryear{Tsujikawa}{Tsujikawa}{2013}]{tsujikawa_quintessence:_2013}
Tsujikawa S.,  2013, \mn@doi [Classical and Quantum Gravity]
  {10.1088/0264-9381/30/21/214003}, 30, 214003

\bibitem[\protect\citeauthoryear{Wolz, Kilbinger, Weller  \& Giannantonio}{Wolz
  et~al.}{2012}]{wolz_validity_2012}
Wolz L.,  Kilbinger M.,  Weller J.,   Giannantonio T.,  2012, \mn@doi [JCAP]
  {10.1088/1475-7516/2012/09/009}, 9, 9

\bibitem[\protect\citeauthoryear{van Erven \& Harremo{\"e}s}{van Erven \&
  Harremo{\"e}s}{2014}]{van_erven_renyi_2014}
van Erven T.,  Harremo{\"e}s P.,  2014, \mn@doi [Transactions on Information
  Theory] {10.1109/TIT.2014.2320500}, 60, 3797

\makeatother
\end{thebibliography}

\appendix

\label{lastpage}
\end{document}